\documentclass[12pt]{article}
\usepackage{indentfirst,graphicx,epsfig}
\usepackage{cite}
\def\eqn{\begin{equation}}
\def\enn#1{\label{#1} \end{equation}}
\def\eq{\[}
\def\en{\]}
\def\r#1{(\ref{#1})}
\def\e#1{eq. (\ref{#1})}
\def\es#1{eqs (\ref{#11}) and (\ref{#12})}

\def\f#1{Figure {\ref{#1}}}
\def\t#1{Table \ref{#1}}

\begin{document}

\title{Projecting to a Slow Manifold: \\
Singularly Perturbed Systems and Legacy Codes.}

\author{C. W. Gear$^{1,2}$
\footnote{
Corresponding author{\tt wgear@princeton.edu, tel (609) 737-7023,
FAX (609) 258-0211 }}, T.J. Kaper${}^{3}$,\\
I.G. Kevrekidis${}^{1,4}$, and A. Zagaris${}^{3}$\\
\small ${}^1$ Department of Chemical Engineering,
Princeton University, Princeton, NJ 08544; \\
\small ${}^2$ NEC Laboratories USA, retired;\\
\small ${}^{3}$ Dept of Mathematics and Center for BioDynamics,
Boston University, Boston, MA 02215;\\
\small ${}^4$ Program in Applied and Computational Mathematics,
Princeton University, Princeton, NJ 08544. \\
}

\date{\today}
\maketitle

\abstract{
The long-term dynamics of many dynamical systems
evolve on an attracting, invariant ``slow manifold"
that can be parameterized by a few observable variables.
Yet a simulation using the full model
of the problem requires initial values for all variables.
Given a set of values for the observables parameterizing
the slow manifold, one needs a procedure for
finding the additional values such that the state is  close to the
slow manifold to some desired accuracy.
We consider problems whose solution
has a singular perturbation expansion, although we do not know what it
is nor have any way to compute it.
We show in this paper that, under some conditions, computing the values of the
remaining variables so that their $(m+1)$st time derivatives are zero
provides an estimate of the unknown variables that is an $m$th-order
approximation to a point on the slow manifold in sense to be defined.
We then show how this criterion can be applied approximately when the system
is defined by a legacy code rather than directly through
closed form equations.}


\newpage

\section{Introduction}

The derivation of reduced dynamic models for many chemical and physical
processes hinges on the existence of a low-dimensional, attracting,
invariant ``slow manifold" characterizing the long-term process dynamics.
This manifold is parameterized by a small number of system variables
(or ``observables", functions of the system variables):
when the dynamics
have approached this manifold, knowing these observables suffices to
approximate the full system state.
A reduced dynamic model for the evolution of the observables
can then in principle be deduced; the resulting simplification
in complexity and in size can be vital in understanding and modeling
the full system behavior.
This reduction has been the subject of intense study from the
theoretical, practical and computational points of view.
Low-dimensional center-unstable manifolds are crucial in the
study of normal forms and bifurcations in dynamical systems
(e.g. \cite{GuckHolmes}); the theory of Inertial Manifolds and
Approximate Inertial Manifolds \cite{Constantin,Temam}
has guided model reduction in dissipative Partial Differential Equations;
the study of fast/slow dynamics in systems of Ordinary Differential
Equations is the subject of geometric singular perturbation theory
(e.g. \cite{Fenichel}).
On the modeling side, the Bodenstein ``quasi-steady state" approximation
has long been the basis for the reduction of complex chemical
mechanisms, described by large sets of ODEs (\cite{Bodenstein},
see also the discussion by Turanyi \cite{Turanyi}).
More recently, an array of computational approaches have been proposed
that bridge singular perturbation theory with large scale scientific
computation for such problems; they include the Computational Singular
Perturbation (CSP) approach of Lam and Goussis \cite{lamg,LG2,LG3},
and the Intrinsic Low-Dimensional Manifold approach of Maas and Pope \cite{MaasPope}.
The mathematical underpinnings of these methods have also been
studied (\cite{Kaper1,Kaper2}).
Lower-dimensional manifolds arise naturally also in the context of
differential-algebraic equations,
where the initialization problem has attracted considerable attention (e.g. \cite{Linda}).

Remarkably, the same concept of separation of time scales and low-dimensional
long-term dynamics underpins the derivation of ``effectively simple"
descriptions of complex systems.
In this context, the detailed model is a collection of agents (molecules,
cells, individuals) interacting with each other and their environment;
the entire distribution of these agents that evolves through atomistic
or stochastic dynamic rules.
In many problems of practical interest it is possible to write
macroscopic equations for the large-scale, coarse-grained dynamics
of the evolving distribution in terms of only a few quantities,
typically lower moments of the evolving distribution.
In the case
of isothermal Newtonian flow, for example, we can write closed
evolution equations for the density and the momentum fields, the zeroth
and first moments of the particle distribution over velocities.
This is again a singularly perturbed problem; only in this case the
higher moments of the evolving distribution have become
quickly slaved to the lower ones (in this case stresses, after
a few collisions, have become functionals of velocity gradients).
Newton's law of viscosity therefore represents a similar type of
``slow manifold" as in the ODE case discussed above - fast variables
(stresses) become functionals of velocity gradients, and the slow
manifold is embodied in the {\it closure}: Newton's law of viscosity.
The use of slow manifolds in non-equilibrium thermodynamics, and more
generally in the study of complex systems, is also a subject of
intense current research (see the work of Gorban, Karlin, Oettinger
and coworkers
\cite{Gorban1,Gorban2}, as well as \cite{Yannis1,Yannis2}) for
some recent computational studies.
In this context, one has an ``inner simulator" at the microscopic or
stochastic level for evolving the detailed distributions; a separation
of time scales does not arise at this level, but rather at the level
of the evolution of the {\it statistics} or {\it moments} of these
distributions.
Typically the lower moments of the evolving distributions
parameterize the slow manifold, while the higher moments quickly
evolve to functionals of the lower ones.
Since we do not have explicit formulas for the equations at the
coarse-grained, macroscopic level, the following interesting question
arises: can we benefit from singular perturbation, when no
closed form evolution equations are available, and the only tool
at our disposal is a ``black box" dynamic simulator of the detailed problem ?
This is the problem we will address in this paper.

We will assume that we are given an evolutionary system which can
be described by
\begin{eqnarray}
u' & = & p(u,v) \label{eq-uv1} \\
v' & = & q(u,v) \label{eq-uv2}
\end{eqnarray}
where prime designates differentiation w.r.t. $t$, the dimensions of
$u$ and $v$ are $N_u$ and $N_v$ respectively, and values, $u(0)$,
are specified only for $u$.
In the case of a legacy dynamic code, we may not even be given the
formulas for these equations explicitly; instead, we may be given a
time-stepper of the above system as a black-box subroutine: a code
that, provided an initial condition for $u$ and $v$, will return an
accurate approximation of $u$ and $v$ after a time interval
(a reporting horizon) $\Delta T$.
We wish to
find $v(0)$ so that the solution is ``close to a slow manifold.''
This
statement is deliberately vague because in practice we are proceeding
on the belief that there exists a slow, attracting, invariant manifold that can be
parameterized  by $u$
and that the variables $v$, in some sense, ``contain'' the fast
variables so that
their values {\it on the slow manifold} can be computed from the values of
$u$ at any time.  (Note that we do not need to {\em know} which are the
slow variables, only to be able to identify a set of variables
sufficient to parameterize the slow manifold.)
This implies that the manifold is the graph of a function $v = v(u)$ over
the observables $u$.
As we proceed, we will make these statements more precise.
However, we
will not cast them into the form of theorems because, even though this is
possible, the conditions for the application of the theorems would be
essentially impossible to verify in all but trivial problems.
As with
many numerical methods, the primary test of applicability is in the
actual application.

We assume that the system can be expressed in terms of other
variables, $x$ and $y$, of the same dimension as $u$ and $v$
respectively, and that in terms of $x$ and $y$ the system can be
written in the usual singular perturbation form:
\begin{eqnarray}
x' & = & f(x,y)\label{eq-sp1} \\
\epsilon y' & = & g(x,y) \label{eq-sp2}
\end{eqnarray}
where $x$ and $y$ are also of dimension $N_u$ and $N_v$,
respectively.
We will assume that their initial values are specified
as $x(0)$ and $y(0)$, independent of $\epsilon$.
The singular perturbation parameter, $\epsilon$, is
associated with the gap, or ratio, between the ``active'' (slow) eigenvalues
and the rightmost of the negative, inactive eigenvalues of the
linearized problem, locally.
We stress that we {\em do not} assume that we know how to express the
equations in the form of \es{eq-sp} nor do we
have any knowledge of the transformation
\begin{eqnarray}
u & = & u(x,y),\hspace{0.1in}x(0,\epsilon) = x(0) \label{eq-tr1} \\
v & = & v(x,y),\hspace{0.1in}y(0,\epsilon) = y(0) \label{eq-tr2}
\end{eqnarray}
other than that we assume that it is well-conditioned and does not have large
derivatives.
We will consider
singular perturbation expansions in $\epsilon$, even though the
parameter is not identified (and cannot be varied). The functions
$f$ and $g$ could also involve the parameter $\epsilon$, but that
serves little in this presentation other than to complicate the
algebra.

The standard singular perturbation expansion for the solution of
\es{eq-sp} takes the form:
\begin{eqnarray}
x(t,\epsilon) & = & \sum_{n=0}^\infty \epsilon^nX_n(t) +
\epsilon\sum_{n=0}^\infty \epsilon^n\xi_n(\tau) \label{sp-sol1} \\
y(t,\epsilon) & = & \sum_{n=0}^\infty \epsilon^nY_n(t) +
\sum_{n=0}^\infty \epsilon^n\eta_n(\tau)  \label{sp-sol2}
\end{eqnarray}
This involves an outer solution $(X(t),Y(t))$, that is smooth in the
sense that its time derivatives are
modest, and an inner solution, $(\xi(\tau),\eta(\tau))$, that captures
the fast boundary layer where the solution
typically changes like $e^{-t/\epsilon} = e^{-\tau}$.
Both outer and inner solutions are expressed as power series in
$\epsilon$, the latter as a function of the {\em stretched} time
$\tau$.
The inner solution is fast in the sense that each
differentiation by $t$ introduces a multiplier of $1/\epsilon$.

We define the slow manifold as the manifold that contains all solutions
of \es{eq-sp} of the form \es{sp-sol} with the inner solution
identically zero.
This is an invariant manifold.
We say that {\it a solution} is an $m$th-order approximation to a slow manifold solution
if it has the form in
\es{sp-sol} with the first $m+1$ terms of the inner solution expansion
identically zero.
A {\it point} $(u,v)$ is an $m$-th order approximation to
the slow manifold, or $m$-th order close to the slow manifold,  if it
lies on an $m$-th order approximate solution.

We want to stress that we are not proposing a technique for finding
the singular perturbation expansion.
Rather, we are using the ideas
as a scaffold for the theoretical justification of the proposed
computational method.
It is possible that the method will provide answers even when
the singular perturbation expansions do not converge (although in that
case we have no justification other than an intuitive one).
The procedure we propose for finding the $v$ values that are close to the slow manifold
given the $u$ values is to find values of $v$ that approximately
solves the $N_v$ dimensional non-linear equation
\eqn
\frac{d^{m+1}v}{dt^{m+1}} = 0
\enn{cond}
which we call ``the [$(m+1)$-st] derivative condition.''
(Compare this with the
``bounded derivative principle'' \cite{Kreiss} which requires the
first $m$ time derivatives to be of order 1.  That condition can be
applied to problems with fast oscillating solutions.   Ours can not,
but it is simpler to apply to the types of problems we consider.)

Note that \e{cond} is a {\em local condition} - that is, it is applied at
a single time - which we will take to be $t = t_0 = 0$ - to determine a
value of $v$ corresponding to a given value of $u$.
A solution of \es{eq-uv} starting from these values of $u$ and $v$ will
not satisfy \e{cond} for $t > 0$ - but we do expect the solution to be
close to the slow manifold.
Intuitively the condition in \e{cond} finds a point close to the slow manifold because
differentiation ``amplifies'' rapidly varying components more than
slowly varying components, so \e{cond} seeks a region where the fast
components are small.
We will suggest ways in
which \e{cond} can be solved approximately in practical codes, even
those based on a legacy code for the integration of \es{eq-uv}.

While the approach will be presented and implemented for
singularly perturbed sets of ODEs, the ``legacy code" formulation
is appropriate also in cases where the inner simulator is not
a differential equation solver, but rather a microscopic / stochastic
simulator.
In the {\it equation free approach} we have been developing for the
computer-assisted study of certain classes of complex systems, the
variables $u$ correspond to macroscopic observables of a microscopic
simulation (typically, moments of a stochastically or deterministically
evolving distribution).
In this case $v$ corresponds to statistics of the evolving distribution
(e.g. higher moments) that become quickly slaved to (become functionals
of) the observables $u$; thus the analogy of a slow manifold persists in
moments space for the evolving distribution.

The paper is organized as follows:
In the next Section we will outline the theoretical basis for the
derivative condition.
In Section 3 we discuss ways in which the derivative condition can
be approximately applied as a difference condition and used with
legacy codes.
Section 4 presents some simple examples of its application.
We conclude with a brief summary and outline of the scope of the method and
some of the challenges we expect to arise in its wider application.

\section{Theoretical Basis}

In this section we will show that the application of the condition
in \e{cond} will lead to a $m$th-order approximation to the slow
manifold under suitable smoothness and smallness conditions.
We
will start by sketching the parts of singular perturbation
expansion theory that we need by paraphrasing the presentation in O'Malley's
monograph\cite{omalley}, particularly pp 46-52.
We will use the
same notation to make it easier for the reader who wishes to get
more detail from that book.

In the following we will write $X_n,~Y_n,~\xi_n$, and $\eta_n$ to mean
$X_n(t),~Y_n(t),~\xi_n(\tau)$, and $\eta_n(\tau)$.
We recall from
\cite{omalley} that the $t$ and $\tau$ dependencies are treated
separately, and that the terms in the outer expansion,
$\{X_n,Y_n\}$, are obtained term by term by substituting \es{sp-sol}
into \es{eq-sp}, and  equating each outer term in $\epsilon^n$ to
zero, starting with $n = 0$.
For $n = 0$ we obtain the DAE:
\begin{eqnarray}
X'_0 & = & f(X_0,Y_0), \hspace{0.1in}X_0(0) = x(0)\label{eq-out01} \\
0 & = & g(X_0,Y_0) \label{eq-out02}
\end{eqnarray}
The existence of a smooth solution of this equation for any $x(0)$
requires the assumption that $g_y$ is non-singular.
(The existence of asymptotic expansions for the inner and outer
components requires the stronger assumption that $g_y$ is a stable
matrix.)
Then, $Y_0$
is specified uniquely in terms of $X_0$ by \e{eq-out02}, say as
\eq
Y_0 = \phi(X_0).
\en

The $n$th term in the power series yields the DAE
\begin{eqnarray}
X'_n & = & f_x(X_0,Y_0)X_n + f_y(X_0,Y_0)Y_n + \tilde{f}_n \label{eq-outn1} \\
0 & = & g_x(X_0,Y_0)X_n + g_y(X_0,Y_0)Y_n + \tilde{g}_n
\label{eq-outn2}
\end{eqnarray}
where the $\tilde{f}_n$ and $\tilde{g}_n$ are defined in terms of
earlier terms in the outer expansion, $\{X_j,Y_j\}, j = 0, \cdots,
n-1$.
The initial condition, $X_j(0)$, has to be specified.  Since we
have set  $X(0) = x(0)$, $X_j(0)$ is obtained from \e{sp-sol1} by
requiring that the $\epsilon^j$ term vanishes at $t = 0$, which
gives
\eqn
X_j(0) = -\xi_{j-1}(0).
\enn{initX}

We are most interested in the way in which the inner terms are
defined, since we wish to annihilate the first $m+1$ of these to get
an $m$th-order approximation.
These are
obtained by considering the change in the inner terms as $\tau$ varies for
arbitrarily small $\epsilon$, in other words, with $t = 0$ and the outer
solutions fixed at their initial values.
Following \cite{omalley}
we consider terms in successive powers of $\epsilon$ and find that the $\epsilon^0$ terms satisfy:
\begin{eqnarray}
\dot{\xi}_0 & = & f(x(0),Y_0(0) + \eta_0) - f(x(0),Y_0(0)) \label{eq-in01} \\
\dot{\eta}_0 & = & g(x(0),Y_0(0) +\eta_0) - g(x(0),Y_0(0)) \label{eq-in02}
\end{eqnarray}
where a dot represents differentiation w.r.t. $\tau = t/\epsilon$.
If
we have an initial value for $\eta_0(0)$ we can solve \e{eq-in02} for
$\eta_0$.
Eq. \r{eq-in01} gives $\xi_0$ as an indefinite integral so
it is determined by specifying $\xi_0$ at any point.
This is normally
done at $\tau = \infty$, in other words, at the end of the boundary
layer.
However, in our development here we will be showing that
$\eta_j$ and $\xi_j$ are identically zero for $j \le m$, so we will
actually choose $\xi_j(0) = 0$ (so that we also have $X_{j+1}(0) = 0$
from \e{initX}).
Subsequent inner terms satisfy
\begin{eqnarray}
\dot{\xi}_n & = & f_y(x(0),Y_0(0) + \eta_0)\eta_n + \hat{f}_n \label{eq-inn1} \\
\dot{\eta}_n & = & g_y(x(0),Y_0(0) +\eta_0)\eta_n + \hat{g}_n \label{eq-inn2}
\end{eqnarray}
where $\hat{f}_n$ and $\hat{g}_n$ are functions of the earlier terms,
$\{\xi_j,\eta_j\}, j = 0, \cdots, n-1$.  In particular, if all of
these terms are zero, then $\hat{f}_n$ and $\hat{g}_n$ are zero.
Eq. \r{eq-inn2} can be solved if an initial value is known for
$\eta_n(0)$.  Once again, \e{eq-inn1} gives $\xi_n$ as an indefinite
integral.

In the following we are going to show, one by one, that $\eta_j(0) = 0$ for $j = 0,
1, \cdots, m$, and that we can then choose $\xi_j(0) = 0$.
Note that once we have shown that $\eta_0(0) = 0$ then \es{eq-in0} indicate that
$\eta_0(\tau) = 0$ and that $\xi_0(\tau)$ is constant, which we can
make zero by choosing $\xi_0(0) = 0$.
Then it follows that $\hat{f}_1$ and $\hat{g}_1$ are
identically zero.  Repeating this argument, we see that if  $\eta_j(0) = 0, \xi_j(0)
= 0, j =  0, \cdots, m$ then $\xi_j(\tau)$ and $\eta_j(\tau)$ are
identically zero for $j \le m$. This provides a solution that
is an $m$th-order approximation.

Now we return to the original problem phrased in terms of $u$ and
$v$.
If we knew the transformation to $x$ and $y$ we would do better
to work in that space, but our assumption is that, although a
transformation exists, it is unknown and we have to work with $u$ and $v$.
We want to show that if the $(m+1)$st derivative of $v$ is zero, then
the point is $m$th-order close to the slow manifold.
All terms below are evaluated
at $t = 0$ or $\tau = 0$ - the time at which we are attempting to
solve \e{cond}.
We will simplify the notation and write $\eta_j$ for
$\eta_j(0)$ and similarly for other terms in the following.
We have from \e{eq-tr2}
\eqn
\frac{d^{m+1}v}{dt^{m+1}} =  v_y\frac{d^{m+1}y}{dt^{m+1}} + {\rm other~terms}
\enn{vdiff}
where the other terms involve either partial derivatives of $v$ w.r.t. $x$
and/or multiple derivatives of $v$ w.r.t. $y$ and products of
derivatives of $y$.

Substituting from \e{sp-sol2} into \e{vdiff} and extracting the lowest
order term in $\epsilon$ ($\epsilon^{-m-1}$) we find that at $t = 0$
\eqn
\frac{d^{m+1}v}{dt^{m+1}} =
\epsilon^{-m-1}v_y\frac{d^{m+1}\eta_0}{d\tau^{m+1}} + {\rm
other~terms} + {\rm O}(\epsilon^{-m})
\enn{exp1}
where now the other terms include products of a higher-order partial derivative of
$v$ w.r.t. $y$ with more than one $\tau$-derivative of $\eta_0$ such that the sum of the
levels of differentiation is $m+1$, that is, terms like
\eq
v_{yy}\frac{d^k\eta_0}{d\tau^k}\frac{d^{m+1-k}\eta_0}{d\tau^{m+1-k}}
\en
and terms with higher partial derivatives and more derivatives of $\eta_0$ in the product.
Note that whenever $\eta_0$ appears,
it is always differentiated w.r.t.
$\tau$ at least once.
Also note that we do not get any terms involving
$\xi_0$ because of the additional $\epsilon$ appearing in front of the
inner solution expansion for $x$ in \e{sp-sol1}.

Now we use \e{eq-in02} to find the higher-order derivatives of
$\eta_0$ w.r.t. $\tau$.
We get for $p > 1$
\eqn
\frac{d^p\eta_0}{d\tau^p} = g^{p-1}_y\dot{\eta}_0 + {\rm
other~terms}
\enn{tauderv}
where the other terms involve
$\dot{\eta}^j_0$ with $j > 1$. Substituting \e{tauderv} in
\e{exp1} we arrive at
\eqn
\frac{d^{m+1}v}{dt^{m+1}} =  \epsilon^{-m-1}\left [v_y
g^{m}_y\dot{\eta}_0 + \sum_{j=2}^{m+1} v_zg_z\dot{\eta}^j_0\right] + {\rm
O(}\epsilon^{-m})
\enn{main}
where the notation $v_zg_z$ stands
for sums of products of various partial derivatives of $v$ and $g$.
Equating the leading term of the right-hand side of \e{main} to
zero, we now have a polynomial equation for $\dot{\eta}_0$ as
\eqn
v_yg^{m}_y\dot{\eta}_0 + \sum_{j=2}^{m+1} v_zg_z\dot{\eta}^j_0 = 0
\enn{poly}
One solution of this is \eq \dot{\eta}_0 = 0 \en and it is an isolated root as
long as $v_yg_y$ is non-singular.
Since we have assumed that $g_y$ is a
stable matrix for the existence of a singularly perturbed solution
(and hence a slow manifold) and that $v$ in some sense spans the fast variables
(meaning that $v_y$ is non singular), this is no problem.
If all other
partial derivatives involved in \e{poly} are ``of order one'' then other
solutions are also of order one - i.e., well separated
from the zero solution.
We will delay discussion of how to avoid
the ``wrong'' solutions for the moment, and assume that we find
the zero solution.
(If the problem is linear, these other terms are
null, so there are no other solutions, and it is only in the case
of high non-linearity when the partial derivatives are large that
these other solutions can become small and cause problems.)

If $\dot{\eta}_0 = 0$ then \e{eq-in02} tells us that $\eta_0 = 0$
because we have assumed that $g_y$ is a stable matrix (in the domain
of interest).
This immediately implies that $\xi_0 = 0$ (or is
a constant that can be absorbed into the outer solution, thus making
$\xi_0 = 0$).

As discussed following \e{eq-inn2}, the vanishing of $\eta_0$ and
$\xi_0$ means that the last terms of \es{eq-inn} are zero for $n = 1$,
making them look similar to \es{eq-in0}.
Therefore the above
argument can now be applied to show that $\eta_1$ and $\xi_1$ are
zero.
This argument can be repeated for higher-order terms
as long as the power of $\epsilon$ in \e{main} remains negative,
in other words, until we have shown that
\eq
\eta_j = \xi_j = 0, \hspace{0.5in} j = 0, \cdots, m
\en

Note that when we have made the $(m+1)$st derivative zero, the lower
order derivatives will not be zero, or even small.
This is
because a small movement away from the slow manifold can make
large changes in the derivatives of the inner solution.
However, the
difference between successive $v$ values as we make $m$ successively larger
is small - of
order $\epsilon^{m+1}$ as we go from the $m$-th to the $m+1$-st
derivative condition as the $v$ values are converging to the slow
manifold.
Hence one way to
solve for zeros of high-order derivatives would be to start by finding
the zeros of the first order derivative, then repeating for
successively higher-order derivatives, each step requiring smaller and smaller
changes to $v$, until we have found the zeros of the $(m+1)$st derivatives
using whatever computational process is appropriate.
(The
computational process is addressed in the next section.) This
procedure helps address the issue of finding the smallest of multiple
roots of \e{poly} since, for $m = 0$, there is only one root so that
the iteration for $m = 1$ and larger $m$ starts with a good
approximation.  If the zero root is well separated from the others, we
will converge to it.

\section{Practical Implementation}

It is often not practical to work with higher-order derivatives of
a differential equation, either because they are algebraically complicated or
because the equations are defined by a ``legacy code'' - that is, as an
implementation of a step-by-step integrator that effectively
cannot be change or analyzed.
(The same would be true if part of
the derivative calculations involved table look-up functions that
could be difficult to differentiate.)
Therefore, we are interested in
methods that do not require direct access to the mathematical
functions constituting the differential equation.
The same rationale applies when the ``inner simulator" simulates
the system at a different level (i.e. in the form of an evolving
microscopic or stochastic distribution).
In this case we only have a time-$T$ map for the macroscopic
observables, that results from running the microscopic simulator
and monitoring the evolution of the observables (e.g. particle densities)
in time \cite{PNAS,GKT02}.

The obvious alternative is to use a forward difference approximation to
the derivative, noting that if
\eqn
\Delta^{m+1} v(t) \equiv \Delta^{m} v(t+H) - \Delta^{m} v(t)
\enn{fdiff}
with $\Delta^0 v(t) = v(t)$, then
\eq
\Delta^{m+1} v(t) = H^{m+1}\frac{d^{m+1}v}{dt^{m+1}} + {\rm O}(H^{m+2})
\en

It turns out that there is a straightforward
way to implement a functional iteration to find a zero
of the $(m+1)$-st forward difference, even if we only have a
`` black box'' code that integrates  \es{eq-uv}.
Suppose we have operators, $\phi$ and $\theta$, that, given values of
$u(t)$ and $v(t)$, yield
approximations to $u(t+H)$ and $v(t+H)$, namely
\eqn
u(t+H) \approx \phi(u(t),v(t))
\enn{bbox1}
\eqn
v(t+H) \approx \theta(u(t),v(t))
\enn{bbox2}
Letting $t_n = t_0 +nH$ and applying \es{bbox} $m+1$ times starting
from $t = t_0$ we can compute approximations
\eqn
u_{j+1} = \phi(u_{j},v_{j}) \approx u(t_{j+1})
\enn{bboxx1}
\eqn
v_{j+1} = \theta(u_{j},v_{j}) \approx v(t_{j+1})
\enn{bboxx2}
for $j = 0, \cdots, m$.
The
functional iteration to find a $v_0$ for a given $u_0$ such that the
$(m+1)$-st difference is zero consists of the following steps:
\begin{enumerate}
\item Start with a given $u_0$ and a guess of $v_0$.
\item Set the iteration number $p = 1$.
\item Set the current iterate $v_0^{(1)} = v_0$
\item Apply \es{bboxx} $m+1$ times starting from $u_0, v_0^{(p)}$ to
generate $v_1^{(p)}, v_2^{(p)}, \cdots, v_{m+1}^{(p)}$.
\item Compute $\delta = (-1)^{m}\Delta^{m+1} v_0^{(p)}$
\item If $\delta$ is small, the iteration has converged.
\item If $\delta$ is not small,
\eqn
v_0^{(p+1)} = v_0^{(p)} + \delta
\enn{viter}
\item Increment $p$ and return to step 4.
\end{enumerate}
If the black box integrator provides a good integration (that is, it
does not introduce spurious oscillations due to near instability) and
$H$ is small enough, this process will converge on a zero of the
difference.

As an illustration we consider the case $m = 0$ and assume that
the integrator is simply forward Euler with a step size such that its
product with the magnitude of the largest eigenvalue is less than
one.
We see that the process for $v$ consists of computing
\eq
\delta = v_1 - v_0 \approx   H\frac{dv}{dt}(t_0).
\en
If this is
insufficiently small, we replace $v_0$ with  $v_0 + \delta = v_1$.
This is just the stationary projection process used in
\cite{Gear2003} and is related to the ``reverse time"
projective integration method in \cite{gkpast}.
In the case of $m = 1$ we compute $\delta$ as
\eq
\delta = -v_2+2v_1 - v_0
\en
If this is insufficiently small, we add it to $v_0$ to get a new
$v_0$ given by
\eq
v_0 = 2v_1 - v_2
\en
This is precisely the
linear interpolant through $v_1$ and $v_2$ back to the starting
point.   It is illustrated in \f{Fderv}.  (This Figure may be a little
confusing because it is drawn in the $u-v$ plane to emphasize that $u$
is being held constant from iteration to iteration.  The backwards
interpolant, however, is really
taking place in the $t-v$ plane.)
\begin{figure}[p]
\centerline{\psfig{figure=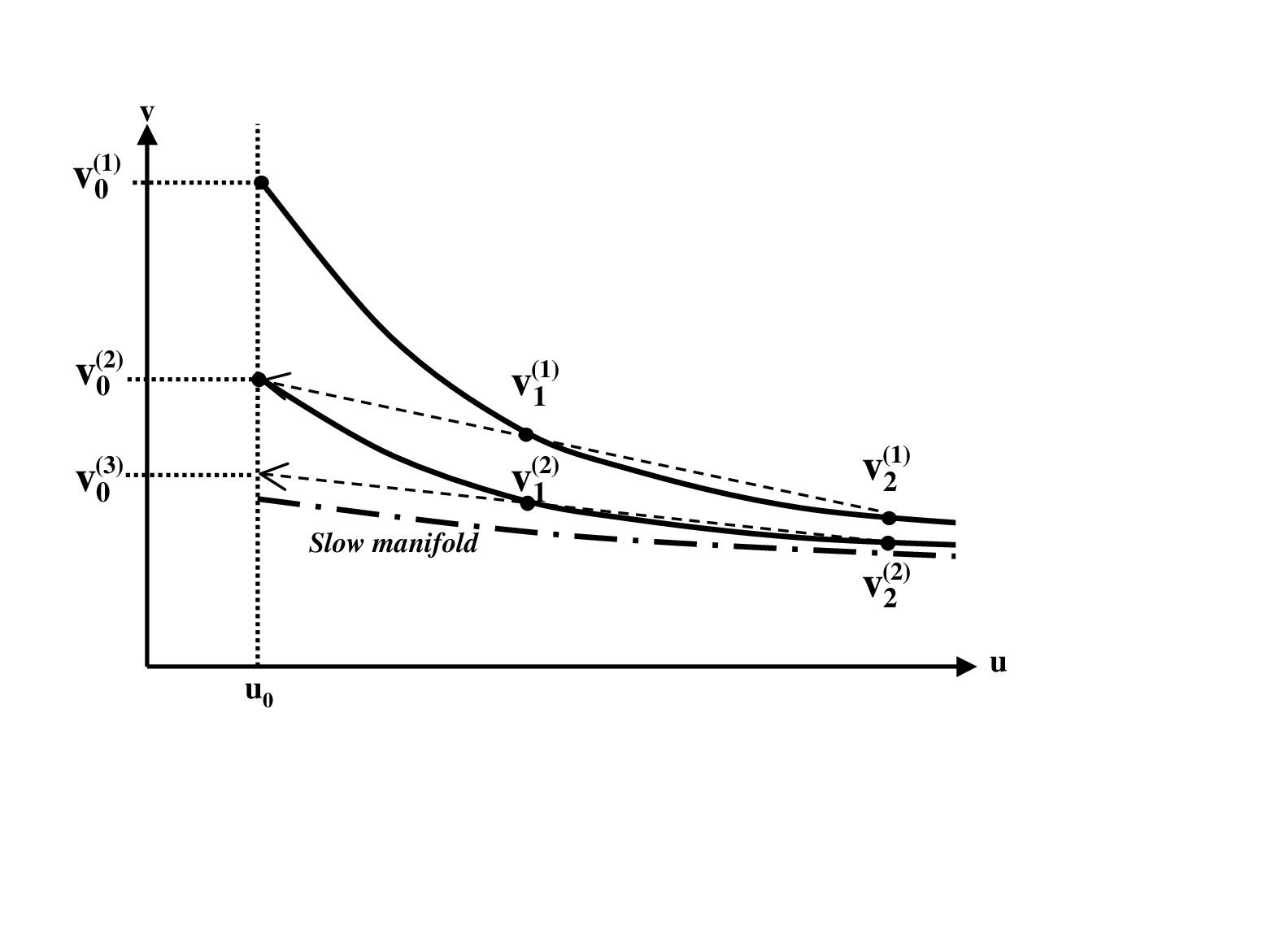,height=3.5in,height=3.5in}}
\vspace{0.1in} \caption{Iterative process for $m = 1$.}\label{Fderv}
\end{figure}
The general iteration is equivalent to the obvious
extension of that: an $m$th degree interpolant is passed through
$v_1, v_2, \cdots, v_{m+1}$ to compute a new approximation to
$v_0$.  This is conveniently done using differences as described
above.

Why does this iteration \e{viter} converge for small $H$ under
reasonable circumstances?
%
Intuitively, the forward integration exponentially damps the fast components more
than they are amplified by the polynomial extrapolation backwards.
In more mathematical terms, the iteration takes the form
\eqn
v^{new}_0 = v_0 + \delta
\enn{iter}
If $\partial \delta/\partial v_0$ is negative definite and small, this
will converge.  We have
\eq
\partial \delta/\partial v_0 = -(-H)^{m+1}\frac{\partial
(\frac{d^{m+1}v}{dt^{m+1}})}{\partial v} + {\rm O}(H^{m+2})
\en
The term
\eq
\frac{\partial(\frac{d^{m+1}v}{dt^{m+1}})}{\partial v}
\en
is dominated by
\eq
\epsilon^{m+1}g_y^{m+1}
\en
in powers of $\epsilon$.  Since we have assumed that $g_y$ is a
negative definite matrix for the existence of a singular perturbation
expansion, convergence follows for small enough $H$.

In the above discussion we have used the attractivity of the manifold
and, in effect, successive substitution in order to converge  to
a fixed point of our mapping.
One can accelerate this computation by using fixed point algorithms,
like Newton's method; clearly, since no equations and Jacobians are
available, the problem lends itself to matrix-free fixed point
implementations like the Recursive Projection Method by Shroff and
Keller \cite{ShroffKeller}
or Newton-Krylov implementations (see \cite{KelleyBook} and, for
a GMRES-based implementation for timesteppers \cite{KKL04}).

\section{Examples}
We will consider three examples to illustrate the  method.  The
Michaelis-Menten enzyme kinetics model is a classic example for singular
perturbation (see \cite{Aris} for a brief discussion on the
history of the model and its analysis).
Since it is a simple system we can make a direct
comparison with an easily computable singular perturbation expansion.
The second example is a realistic five-dimensional chemical reaction
system with a one dimensional slow manifold.
As in
many real examples, we do not know the slow manifold, and can only
show ``plausibility'' of our solution.
The final example is a
contrived five-dimensional non-linear system with a known
two-dimensional slow manifold so that we can compute the ``errors'' as
the distance from the slow manifold.

\subsection{Michaelis-Menten equation}
This is given in a
singular perturbation form in \cite{omalley} as
\begin{eqnarray}
x' & = & -x + (x + \kappa - \lambda)y\label{eq-mm1} \\
\epsilon y' & = & x - (x + \kappa)y \label{eq-mm2}
\end{eqnarray}
with $x(0) = 1$.  For some simple tests we will take $\kappa = 1$,
$\lambda = 0.5$, and $\epsilon =$ 0.1 and 0.01.  (These are larger
figures than typical for reactions, but we wish to show that the method works even
for problems with a relative small gap, and also to have problems
where the higher-order initializations are visibly different from the
lowest order ones.)

The first few terms of the outer solution are
\eqn
y = \frac{x}{x + \kappa} + \frac{\kappa\lambda x}{(x +
\kappa)^4}\epsilon + \frac{\kappa\lambda x(2\kappa\lambda - 3x\lambda
-x\kappa - \kappa^2)}{(x+\kappa)^7}\epsilon^2 +{\rm O}(\epsilon^3).
\enn{expe}

In the following tests, we implemented the operators $\phi$ and
$\theta$ in \es{bbox} using $n$ steps of forward Euler with step size
$h$.  In all cases \e{iter} was iterated until $\delta$ was less than
$10^{-14}$ (which is rather excessive, but we didn't want any errors
from premature termination to color the results).  We ran with $m = 0,
\cdots, 4$ and compared them with the first $m+1$ terms of  \e{expe}
through $m = 2$.  For $\epsilon = 0.1$, the results are shown in
\t{T1} with $h = \epsilon/100$ and $n = 1$, and in \t{T2} with $h =
\epsilon/10$ and $n = 4$.  The tables show the computed approximation to the slow manifold,
$y(0)$, for $x(0) = 1$.  The tables also show the first $m$ terms of
the asymptotic expansion for $m \le 2$.  As can be seen, the discrepancies are
larger when $H = nh$ is larger.

A larger ``integration time horizon'' $H$ gives more rapid convergence.  
However, this means than the difference estimate is a less
accurate approximation to the $(m+1)$-st derivative.  Our theory shows
that making the $(m+1)$-st derivative zero puts the solution on an
$m$-th order approximation to the slow manifold.  Any error in the
derivative estimate creates an error of the same order in the
solution, so we should choose the time horizon so that the errors in
the derivative estimate are of the same order as those we are willing
to tolerate in the approximation to the slow manifold.  This suggests
that an initial approximation could be calculated with a larger $H$
(and small $m$) to get faster convergence,
and then $H$ could be reduced as $m$ is increased to increase the accuracy.

\begin{table}
\begin{center}
\caption{Michaelis-Menten example with $\epsilon = 0.1$, $h = \epsilon/100$ and $n =
1$.}
\begin{tabular}{ccc} \hline
$m$ & Computed $y$ & Asymptotic $y$  \\ \hline
 0 & 0.500000000 & 0.500000000\\
 1 & 0.503049486 & 0.503125000\\
 2 & 0.503031986 & 0.503027344\\
 3 & 0.503031924 & -\\
 4 & 0.503031924 & -\\ \hline
\end{tabular}
\label{T1}
\end{center}
\end{table}

\begin{table}
\begin{center}
\caption{Michaelis-Menten example with $\epsilon = 0.1$, $h = \epsilon/10$ and $n =4$.}
\begin{tabular}{ccc} \hline
$m$ & Computed $y$ & Asymptotic $y$  \\ \hline
0 & 0.498886090 & 0.500000000 \\
1 & 0.503067929 & 0.503125000 \\
2 & 0.503035446 & 0.503027344 \\
3 & 0.503035098 & - \\
4 & 0.503035128 & - \\ \hline
\end{tabular}
\label{T2}
\end{center}
\end{table}

The next case, shown in \t{T3}, uses an $\epsilon$ smaller by a factor
of 10.  Because the higher-order terms in $\epsilon$ are now smaller,
the agreement with the terms in the asymptotic expansion is better.
(However, we are really interested in the agreement of the computed
$v_0$ with the $v$ on the slow manifold, rather than with the
ability to match the first few terms in the asymptotic expansion.)

\begin{table}
\begin{center}
\caption{Michaelis-Menten example with $\epsilon = 0.01$, $h = \epsilon/100$ and $n =1$.}
\begin{tabular}{ccc} \hline
$m$ & Computed $y$ & Asymptotic $y$  \\ \hline
0 & 0.500000000 & 0.500000000 \\
1 & 0.500311725 & 0.500312500 \\
2 & 0.500311533 & 0.500311523 \\ \hline
\end{tabular}
\label{T3}
\end{center}
\end{table}

The cases above ``constrained'' the derivative of the singularly
perturbed ``fast" variable, $y$.  Usually we cannot isolate this variable.
To see the effect of having a different variable set, we transform $x$
and $y$
into
\eqn
u = x + y
\enn{tr-uv1}
\eqn
v = y - x,
\enn{tr-uv2}
and work with $u$ and $v$ assuming that we do not know this
transformation.  (We chose this transformation because in some sense
it puts equal parts of the slow and fast variables, $x$ and $y$, in
$u$ and $v$, illustrating the fact that we need only know variables
that parameterize the slow manifold ($u$ in this case), not ones that
are in some sense dominated by the slow manifold.)  We assumed that we
were given a value of $u$, $u(0) = 1.5$, and computed the
approximation to $v(0)$ using our
method.  This was run with $h = \epsilon/10$ and $n = 4$ for
$\epsilon = 0.1$ in \t{T4} and for $\epsilon = 0.01$ in \t{T5}.  The
tables show the corresponding $x$ and $y$ values derived from $u =
1.5$ and the computed $v(0)$.  The column labeled $ytrue$ gives the
first three terms of the asymptotic expansion of $y$ given the $x$ shown
in the first column. Now we can see the significant error that the $m
= 1$ case gives rise to when we ``come at the slow manifold from a
different angle.''  However, the higher-order approximations yield a
good approximation to the slow manifold.

\begin{table}
\begin{center}
\caption{Michaelis-Menten example with $\epsilon = 0.1$, $h =
\epsilon/10$ and $n = 4$.}
\begin{tabular}{cccc} \hline
$m$ &        $x$  &              $y$ &            $ytrue$ \\ \hline
  0 &      0.98825957 &      0.51174043 &      0.50011008 \\
  1 &      0.99743598 &      0.50256402 &      0.50239316 \\
  2 &      0.99756721 &      0.50243279 &      0.50242566 \\ \hline
\end{tabular}
\label{T4}
\end{center}
\end{table}

\begin{table}
\begin{center}
\caption{Michaelis-Menten example with $\epsilon = 0.01$, $h =
\epsilon/10$ and $n = 4$.}
\begin{tabular}{cccc} \hline
$m$ &        $x$  &              $y$ &            $ytrue$ \\ \hline
  0 &      0.99874363 &      0.50125637 &      0.49999762 \\
  1 &      0.99974927 &      0.50025073 &      0.50024891 \\
  2 &      0.99975069 &      0.50024931 &      0.50024927 \\ \hline
\end{tabular}
\label{T5}
\end{center}
\end{table}

\subsection{A Simplified Hydrogen-Oxygen Reaction System}

We will use an example from Lam and Goussis\cite{lamg}.  It contains
seven radicals, ${\rm O}_2$, H,
OH, O, ${\rm H}_2$, ${\rm H}_2$O, and H${\rm O}_2$ which we will group
in that order as the vector ${\bf y}$.  The differential equations are:
\begin{eqnarray}
\frac{dy_1}{dt} & = &-k_{1f}y_1y_2 + k_{1b}y_3y_4 + k_{4f}y_3y_7 - \mu
k_{5f}y_1y_2 \nonumber \\
\frac{dy_2}{dt} & = &-k_{1f}y_1y_2 + k_{1b}y_3y_4 + k_{2f}y_4y_5 -
k_{2b}y_2y_3 + k_{3f}y_3y_5 - k_{3b}y_2y_6 - \mu k_{5f}y_1y_2 \nonumber \\
\frac{dy_3}{dt} & = & k_{1f}y_1y_2 - k_{1b}y_3y_4 + k_{2f}y_4y_5 +
k_{2b}y_2y_3 \nonumber \\
&& - k_{3f}y_3y_5 + k_{3b}y_2y_6 - k_{4f}y_3y_7
-2k_{8f}y_3^2 + 2k_{8b}y_4y_6 \nonumber \\
\frac{dy_4}{dt} & = & k_{1f}y_1y_2 - k_{1b}y_3y_4 - k_{2f}y_4y_5 +
k_{2b}y_2y_3 + k_{8f}y_3^2 - k_{8b}y_4y_6 \nonumber \\
\frac{dy_5}{dt} & = &-k_{2f}y_4y_5 + k_{2b}y_2y_3 - k_{3f}y_3y_5 +
k_{3b}y_2y_6  \nonumber \\
\frac{dy_6}{dt} & = & k_{3f}y_3y_5 - k_{3b}y_2y_6 +k_{4f}y_3y_7
+ k_{8f}y_3^2 - k_{8b}y_4y_6 \nonumber \\
\frac{dy_7}{dt} & = & -k_{4f}y_3y_7 +\mu k_{5f}y_1y_2  \nonumber
\end{eqnarray}

The values of the coefficients are taken from the cited paper and
shown in \t{Tk}.
The parameter $\mu$ represents pressure and is $4.5 \times 10^{-4}$.
The differential system has two constants of integration representing
mass balance for oxygen and hydrogen atoms, so it is really a
five-dimensional system.  The eigenvalues of a local linearization
in the region of operation for this example are approximately
as shown in \t{Te}.
From these we see that after an interval of order one millisecond, the
system is effectively one dimensional.

\begin{table}
\begin{center}
\caption{Reaction Rates for Hydrogen Oxygen System}
\begin{tabular}{cll} \hline
$i$ &     $k_{if}$  &              $k_{ib}$ \\ \hline
  1 &     $1.0136 \times 10^{12}$ &  $1.1007 \times 10^{13}$ \\
  2 &     $3.5699 \times 10^{12}$ &  $3.2105 \times 10^{12}$ \\
  3 &     $4.7430 \times 10^{12}$ &  $1.8240 \times 10^{11}$ \\
  4 &     $6.0000 \times 10^{13}$ &   \\
  5 &     $6.2868 \times 10^{15}$ &   \\
  8 &     $6.5325 \times 10^{12}$ &  $3.1906 \times 10^{11}$ \\ \hline
\end{tabular}
\label{Tk}
\end{center}
\end{table}

\begin{table}
\begin{center}
\caption{Eigenvalues}
\begin{tabular}{c} \hline
$-2.5 \times 10^{6}$\\
$-1.4 \times 10^{6}$\\
$-4.0 \times 10^{4}$\\
$ -8.3 \times 10^{3}$\\
$-4.0 \times 10^{-3}$ \\ \hline
\end{tabular}
\label{Te}
\end{center}
\end{table}

In the test below we chose $y_5$ (${\rm O}_2$) as the observable
variable.  The system was run from the starting conditions given in
\cite{lamg}
until $t = 6.41 \times 10^{-4}$ (one of the reporting times in their
paper).  Then our process was applied, fixing $y_5$ to its current
value, and choosing all other variables so that their $(m+1)$-st forward
difference is zero.  To emulate a ``legacy code'' situation, we
integrated the equations using a standard integrator ({\bf ode23s} in
MATLAB) over $m+1$ intervals of length $H$. In this example, we used
$H =  1 \times 10^{-5}$. The relative and absolute error tolerances
for {\bf ode23s} were set to $10^{-12}$ and $10^{-14}$, respectively.
To maintain the mass balance relationship,
radical concentrations of H and OH are
computed directly from the mass balance relations.  (These two were chosen
because their concentrations remain reasonably non-zero.  If a
radical whose concentration gets close to zero is used, there is some
danger of roundoff errors causing the concentration to become
negative.    This will often make the system unstable - as well as
physically unrealistic.)

The procedure was run with $m = 0, 1, \cdots, 4$.  The starting values
of the concentrations were as shown in \t{Tinit}.
(These are given for the sake of completeness should anyone wish to
compare with our results.)
The constrained results for each $m$ differ from
the starting value by no more than $10^{-14}$ and from each other by
less, so are not particularly revealing to study directly.  (Larger
changes from the starting value could be obtained by starting from a
different point with the same mass balance values, but would not give
any further insight.)
Since it is difficult to compute the slow manifold (often a
problem when real examples are used) we do not have a good way to
characterize errors, but we can examine two features to see if the method
appears to work.

In \t{Tdiff} we show the differences between the starting value and
the constrained value of $y_4$ for each order of constraint.
We see that these differences show signs of ``converging'' - but this
is certainly not irrefutable evidence of convergence.  As a second
test, we considered the relationship of the local derivative of the
solution at the result of the constraint iteration.  If it were very
far from the slow manifold, we would expect it to have relatively
large components of the eigenvectors corresponding to the large
eigenvalues.  (In general, even on the slow manifold it will not have
zero components in the large eigendirections except for linear
problems.) To estimate the amount of large eigencomponents present we
computed $v = dy/dt = f(y)$ and the local
Jacobian $J = \partial f/\partial y$ at the solution, $y$, of the
constraint iteration.  Then we computed the norm ratio
\eq
R = \frac{||Jv||}{||v||}
\en
Its upper bound is the magnitude of the largest eigenvalue, and a
value significantly less than this is an indicator that $u$
is deficient in the largest eigendirection.  Hence, the norm ratio
gives some indication of the amount of the largest
eigencomponents present.  It values are shown in \t{Trat}.

\begin{table}
\begin{center}
\caption{Radical Concentrations Prior to Constraining to Slow Manifold}
\begin{tabular}{cc} \hline
$y_1$ & $4.2783465727 \times 10^{-13}$ \\
$y_2$ & $3.9878034748 \times 10^{-8}$ \\
$y_3$ & $1.3883748623 \times 10^{-10}$ \\
$y_4$ & $1.1300067412 \times 10^{-11}$ \\
$y_5$ & $4.4019256520 \times 10^{-7}$ \\
$y_6$ & $3.9848995981 \times 10^{-8}$ \\
$y_7$ & $5.3981503775 \times 10^{-15}$ \\ \hline
\end{tabular}
\label{Tinit}
\end{center}
\end{table}

\begin{table}
\begin{center}
\caption{Difference between starting $y_4$ and constrained value}
\begin{tabular}{cc} \hline
$m$ &  Difference \\ \hline
0& $-2.0767748211 \times 10^{-17}$\\
1& $-2.0157837979 \times 10^{-17}$\\
2& $-2.0157711737 \times 10^{-17}$\\
3& $-2.0157697197 \times 10^{-17}$\\
4& $-2.0157429383 \times 10^{-17}$ \\ \hline
\end{tabular}
\label{Tdiff}
\end{center}
\end{table}

\begin{table}
\begin{center}
\caption{Norm Ratio $||Jv||/||v||$ at Constrained Solution}
\begin{tabular}{cc} \hline
$m$ &        $R$ \\ \hline
0 & $4.44973316 \times 10^{5}$\\
1 & $5.85785391 \times 10^{1}$\\
2 & $6.18695075 \times 10^{1}$\\
3 & $2.06227270 \times 10^{2}$\\
4 & $1.50929245 \times 10^{2}$ \\ \hline
\end{tabular}
\label{Trat}
\end{center}
\end{table}

Since the largest eigenvalue is around $2.5 \times 10^6$, it is clear that
the $m = 0$ case contains no more than around 20\% of the large eigendirections, but this
is drastically reduced for $m = 1$.  From this particular starting
point and choice of $H$, larger $m$ gave no further improvement, but
other choices of starting points or $H$ yield norm ratios that reduce with each $m$ increase,
although by relatively small amounts.

\subsection{A Five-Dimensional System}

Because of the difficulty of determining whether the method is getting
better approximations as $m$ increases, our final example is an
artificial non-linear five-dimensional problem with a two-dimensional
attractive invariant manifold.  We start with the loosely coupled differential equations:
\eq
\frac{dx_1 }{dt}= - x_2
\en
\eq
\frac{dx_2 }{dt}=  x_1
\en
\eq
\frac{dw }{dt}= L(x_1^2 + x_2^2 - w)
\en
\eq
\frac{du_1 }{dt}= \beta_1 + u_1^2
\en
\eq
\frac{du_2 }{dt}= \beta_2 + u_2^2
\en
with $L = 1000, \beta_1 = 800, \beta_2 = 1200$.
The solutions of these are
\begin{eqnarray}
x_1 &=& A\cos(t+\phi)\nonumber \\
x_2 &=& A\sin(t+\phi)\nonumber \\
w &=& A(1+be^{-Lt})\nonumber \\
u_i &=& -\beta_i/(1+c_ie^{-\beta_it})\nonumber
\end{eqnarray}
For any starting conditions, $w \rightarrow x_1^2(0) + x_2^2(0)$, and, if
$u_i(0)$ is chosen appropriately, $u_i \rightarrow -\beta_i$ and the
system goes to a closed orbit where the eigenvalues of the system
Jacobian at each point on the closed orbit
are $\pm i$ , -800, -100, and -1200.  Thus $w = x_1^2 + x_2^2$ is an attractive
two dimensional invariant
manifold.  The above system is now subject to the unitary linear
transformation given by $y = Qv$ where $v = [x^T, w, u^T]^T$ and $Q$ is
\eq
Q =\frac{1}{5} \left [ \begin{array}{rrrrr}
-3 & 2& 2& 2& 2\\
2& -3& 2& 2& 2\\
2& 2& -3& 2& 2\\
2& 2& 2& -3& 2\\
2& 2& 2& 2& -3\\ \end{array} \right ]
\en

As in the first example, this is chosen to ``mix up'' the slow and
fast components.
We applied the constraint method using $y_1$ and $y_2$ as the fixed
``observables.''  They were set to the values -791.2 and -792.2
respectively.
The subspace $y_1 = -791.2, y_2 = -792.2$ intersects with the
invariant manifold at four points (the  defining system is a pair of
quadratic equations).  The intersection in the
neighborhood of the solution has the values
\eq
x_1 =-3.559434800714, \hspace{0.2in} x_2 =  -2.559434800714
\en
Integration of $y$ was performed using forward Euler with step size
$h$ for $m+1$ steps, and iterating until the $(m+1)$-st forward
difference was zero, with $m$ = 1, 2, and 3.  The differences between
the constraint calculations and the known solution are shown for
several values of $h$ in \t{T52}.  (They are called ``residuals''
there, since they are not exactly ``errors.'')
We see that the residuals decrease by two to three orders of magnitude
for each increase in $m$ except for larger $h$.  Larger $h$ makes the
iteration converge much more rapidly, but the increased error in the
approximation of the difference to the derivative decreases the degree
of approximation to the slow manifold.  In a practical application, it
might be wise to use a large $h$ to get an initial approximation and
then refine with a smaller $h$, although it might still be wise to use
some convergence acceleration technique.

\begin{table}
\begin{center}
\caption{Residuals in Constraint Solutions.}
\begin{tabular}{r|r|rrrrr} \hline
\multicolumn{1}{c|}{$h$} & \multicolumn{1}{c|}{$m$} & \multicolumn{5}{c}{Residual}
\\ \cline{3-7}
& & \multicolumn{1}{c}{$x_1$} & \multicolumn{1}{c}{$x_2$} & \multicolumn{1}{c}{$w$} & \multicolumn{1}{c}{$u_1$} & \multicolumn{1}{c}{$u_2$}\\ \hline
$8.0 \times 10^{-4}$
& 1 & $-4.84 \times 10^{-4}$ & $-4.84 \times 10^{-4}$ & $3.92 \times 10^{-3}$ & $-2.50 \times 10^{-3}$ & $-1.67 \times 10^{-3}$ \\
& 2 & $-4.34 \times 10^{-6}$ & $-4.34 \times 10^{-6}$ & $2.55 \times 10^{-5}$ & $-1.91 \times 10^{-5}$ & $-8.50 \times 10^{-6}$ \\
& 3 & $-1.21 \times 10^{-6}$ & $-1.21 \times 10^{-6}$ & $-6.08 \times 10^{-7}$ & $3.91 \times 10^{-9}$ &
$1.16 \times 10^{-9}$ \\ \hline
$2.0 \times 10^{-4}$
& 1 & $-4.84 \times 10^{-4}$ & $-4.84 \times 10^{-4}$ & $3.92 \times 10^{-3}$ & $-2.50 \times 10^{-3}$ & $-1.67 \times 10^{-3}$ \\
& 2 & $-3.43 \times 10^{-6}$ & $-3.43 \times 10^{-6}$ & $2.59 \times 10^{-5}$ & $-1.91 \times 10^{-5}$ & $-8.50 \times 10^{-6}$ \\
& 3 & $-3.01 \times 10^{-7}$ & $-3.01 \times 10^{-7}$ & $-1.55 \times 10^{-7}$ & $3.73 \times 10^{-9}$ &
$1.14 \times 10^{-9}$ \\ \hline
$5.0 \times 10^{-5}$
& 1 & $-4.84 \times 10^{-4}$ & $-4.84 \times 10^{-4}$ & $3.91 \times 10^{-3}$ & $-2.49 \times 10^{-3}$ & $-1.66 \times 10^{-3}$ \\
& 2 & $-3.22 \times 10^{-6}$ & $-3.22 \times 10^{-6}$ & $2.64 \times 10^{-5}$ & $-1.95 \times 10^{-5}$ & $-8.54 \times 10^{-6}$ \\
& 3 & $-7.55 \times 10^{-8}$ & $-7.55 \times 10^{-8}$ & $-3.09 \times 10^{-8}$ & $-7.27 \times 10^{-9}$ &
$4.20 \times 10^{-10}$ \\ \hline
\end{tabular}
\label{T52}
\end{center}
\end{table}

\section{Discussion and Conclusion}

We presented a ``computational wrapper" approach for the approximation of a low-dimensional
slow manifold using a legacy simulator.
The approach effectively constitutes a protocol for the design and processing of
brief computational experiments with the legacy simulator, which converge to
an approximation of the slow manifold; in the spirit of CSP, one can think of
it as ``singular perturbation through computational experiments".
It is interesting that, if one could initialize a laboratory experiment at will,
our ``computer experiment" protocol could become a laboratory experiment protocol
for the approximation of a slow manifold.

The approach can be enhanced in many ways; we already mentioned the possible
use of matrix-free fixed point algorithms for the acceleration of its convergence.
Here we used the ``simplest possible" estimation (through finite difference
interpolation) of the trajectory from the results of the simulation.
Better estimation techniques (e.g. maximum likelihood) can be linked with the
data processing part of the approach; this will be particularly important when
the results of the detailed integration are noisy, as will be the case in the
observation of the evolution of statistics of complex evolving distributions.

It is also important to notice that, upon convergence of the procedure, one can
implement a matrix-free, timestepper based computational approximation of the
leading eigenvalues of the local linearization of the dynamics (e.g. through
a timestepper based Arnoldi procedure, see \cite{Christodoulou,Siettos}).
As the evolution progresses, or as the parameters change, this test can be used
to adaptively adjust the local {\it dimension} of the slow manifold - we can
detect whether a slow mode is starting to become fast, or when a mode that used
to be fast is now becoming slow.
The {\it eigenvectors} of these modes constitute good {\it additional observables}
for the parameterization of a ``fatter" slow manifold.
One of the important features of the approach is that one does not need to
{\it a priori} know what the so-called ``slow variables" are - {\it any}
set of observables that can parameterize the slow manifold (i.e., over
which the manifold is the graph of a function) can be used for our approach.
If data analysis (\cite{PCA,SparseKernel}) suggests good observables
that
are nonlinear combinations of the ``obvious" state variables, the
approach can still be implemented; the knowledge of good ``order parameters"
can thus be naturally incorporated in this approach.

Overall, this approach provides us with a {\it good initial condition} for the
full problem, consistent with a set of observables - an initial condition that
lies close to the slow manifold, sometimes referred to as a ``mature" or ``bred"
initial condition.
Such initial conditions are essential for the implementation of equation-free
algorithms: algorithms that solve the reduced problem {\it without} ever
deriving it in closed form \cite{GK1,GK2,GK3}).
Indeed, short bursts of appropriately initialized simulations can be used
to perform long term prediction (projective and coarse projective integration)
for the reduced problem, its stability and bifurcation analysis, as well
as tasks like control and optimization.
We expect this approach to become a vital component of the ``lifting" operator
in equation-free computation.

\section*{Acknowledgments}
This work was partially supported by AFOSR (Dynamics and Control)
and an NSF/ITR grant (C.W.G, I.G.K), and NSF Grant \#0306523, Div. Math
Sci., (T.J.K, A.Z).


\newpage


\begin{thebibliography}{10}

\bibitem{Aris} R. Aris, {\it Mathematical Modeling; A Chemical
Engineer's perspective}, Academic Press, San Diego (1999)

\bibitem{Bodenstein} M. Bodenstein, Eine Theorie der photochemischen
Reaktionsgeschwindigkeiten, {\it Z. Phys. Chem.} {\bf 85} pp.329-397 (1913)

\bibitem{Linda} P. N. Brown, A. C. Hindmarsh and L. R. Petzold, Consistent
initial condition calculation for differential-algebraic systems,
{\it SIAM J. Sci. Comput.}, {\bf 19} 1495. (1998)

\bibitem{GK3}   L. Chen, P. G. Debenedetti, C. W. Gear and I.G.Kevrekidis,
>From molecular dynamics to coarse self-similar solutions: a simple
example using
equation-free computation",  {\it J.Non-Newtonian Fluid Mech.} {\it in press} (2004).

\bibitem{Christodoulou} K. N. Christodoulou and L. E. Scriven,
Finding leading modes of a viscous free surface flow: an asymmetric
generalized eigenproblem.
{\it Quart. Appl. Math.} {\bf 9} 17 (1998)

\bibitem{Constantin} P. Constantin, C. Foias, B. Nicolaenko and R. Temam,
{\it Integral Manifolds and Inertial Manifolds for Dissipative Partial
Differential Equations}, Springer Verlag, NY. (1988)

\bibitem{Fenichel} N. Fenichel, Geometric singular perturbation theory for
ordinary differential equations, {\it J. Diff. Equ.} {\bf 31} pp.53-98 (1979)

\bibitem{Gear2003}
C.~W. Gear and I.~G. Kevrekidis, Constraint-defined manifolds: a legacy-code
  approach to low-dimensional computation, {\it J. Sci. Comp.} in press, (2004);
  also physics/0312094 at arXiv.org.

\bibitem{gkpast}  C. W. Gear and I. G. Kevrekidis, Computing in the Past
with Forward Integration,
{\it Physics Letters A}. {\bf 321} pp.335-343  (2004); also
as nlin.CD/0302055 at arXiv.org

\bibitem{GKT02} C. W. Gear, I.G.Kevrekidis and C. Theodoropoulos,
``Coarse" Integration/Bifurcation Analysis via Microscopic
Simulators: micro-Galerkin methods,
{\it Comp. Chem. Engng.} {\bf 26} pp.941-963 (2002)

\bibitem{Gorban2} A. N. Gorban and I. V. Karlin
Geometry of irreversibility: the film of nonequilibrium states, {\it
Phys. Rep.} in press, (2004); also
cond-mat/0308331 at arXiv.org

\bibitem{Gorban1} A. N. Gorban, I. V. Karlin and A. Yu. Zinovyev,
Constructive methods of invariant manifolds for kinetic problems,
cond-mat/0311017 at arXiv.org

\bibitem{GuckHolmes}
J. Guckenheimer and P. Holmes, {\it Nonlinear Oscillations, Dynamical Systems
and Bifurcations of Vector Fields}, Spinger-Verlag, NY (1983)

\bibitem{PCA}
I.~T. Jolliffe, {\em Principal Component Analysis\/}, Springer-Verlag,
New York, NY (1986)

\bibitem{Kaper1} H. G. Kaper and T. J. Kaper, Asymptotic analysis of two reduction methods
for systems of chemical reactions, {\it Physica D} {\bf 65} pp.66-93 (2002)

\bibitem{KelleyBook}
C.~T. Kelley, Iterative Methods for Linear and Nonlinear Equations,
SIAM Publications, Philadelphia (1995)

\bibitem{KKL04} C. T. Kelley, I. G. Kevrekidis and L. Qiao,
Newton-Krylov solvers for time-steppers, submitted to {\em SIAM Dyn.
Systems}, (2004); also math.DS/0404374 at arXiv.org

\bibitem{Yannis1} I. G. Kevrekidis, C. W. Gear, J. M. Hyman,
P. G. Kevrekidis, O. Runborg and K. Theodoropoulos,
Equation-free coarse-grained multiscale computation: enabling microscopic
simulators to perform system-level tasks,
{\it Comm. Math. Sciences} {\bf 1}(4) pp.715-762 (2003); original version
can be obtained as physics/0209043 at arXiv.org.

\bibitem{Yannis2}
I.~G. Kevrekidis, C.~W. Gear, and G.~Hummer, Equation-Free: The
computer-aided analysis of complex multiscale systems, {\it AIChE Journal}, in
press  (2004).

\bibitem{Kreiss} H-O.~Kreiss, Problems with Different Time Scales, in
Multiple Time Scales, ed. J. H. Brackbill and B. I. Cohen, pp 29-57, Academic
Press (1985)

\bibitem{LG2} S. H. Lam, Using CSP to understand complex chemical kinetics,
{\it Combust. Sci. Technol.} {\bf 89} pp.375-404 (1993)

\bibitem{LG3}S. H. Lam and D. A. Goussis, The CSP method for simplifying
chemical kinetics, {\it Int. J. Chem. Kin.} {\bf 26} pp.461-486 (1994).

\bibitem{lamg} S.~H.~Lam and D.~A.~Goussis, Understanding Complex
Chemical Kinetics with Computational Singular Perturbation,:  22nd
Symposium on Combustion, The Combustion Institute, pp.931-941 (1988)

\bibitem{MaasPope} U. Maas and S. B. Pope, Simplifying chemical kinetics:
intrinsic low-dimensional manifolds in composition space, {\it Comb. Flame}
{\bf 88} pp.239-264 (1992)

\bibitem{GK1} A. G. Makeev, D. Maroudas, A. Z. Panagiotopoulos and I.G.
Kevrekidis,
Coarse bifurcation analysis of kinetic Monte Carlo simulations: a
lattice gas model with lateral interactions, , {\it J. Chem. Phys.} {\bf
117}(18) pp.8229-8240 (2002)

\bibitem{omalley} R.~E.~O'Malley, Singular Perturbation Methods for
Ordinary Differential Equations, Applied Mathematical Scinces Vol 89,
Springer-Verlag, Berlin (1991)

\bibitem{GK2} R. Rico-Martinez, C. W. Gear and I. G. Kevrekidis,
``Coarse Projective KMC Integration: Forward/Reverse Initial and
Boundary Value Problems",
{\it J. Comp. Phys.}, {\bf 196}(2) pp.474-489 (2004)

\bibitem{ShroffKeller}
G.~M. Shroff and H.~B. Keller, Stabilization of unstable procedures: the
recursive projection method, {\it SIAM J. Numer. Anal.}, {\bf 30} pp. 1099--1120
(1993)

\bibitem{Siettos}
C. Siettos, M.~D. Graham, and I.~G. Kevrekidis, Coarse {B}rownian
dynamics for nematic liquid crystals: bifurcation, projective integration and
control via stochastic simulation, {\it J. Chem. Phys.}, {\bf 118} pp.10149--10157
(2003)

\bibitem{SparseKernel}
A.~J. Smola, O.~L. Mangasarian, and B.~Schoelkopf, Sparse
Kernel Feature Analysis, in {\em 24th Annual Conference of Gesselschaft
f{\"u}r Klassifikation\/} (University of Passau, Passau, Germany, 2000), data
Mining Institute Technical Reort 99-04 (1999)

\bibitem{Temam} R. Temam  {\it Infinite Dimensional Dynamical Systems
in Mechanics and Physics} Springer Verlag, NY. (1988)

\bibitem{PNAS} K. Theodoropoulos, Y.-H. Qian and I.G.Kevrekidis,
``Coarse" stability and bifurcation analysis using timesteppers:
a reaction diffusion example,
{\it Proc. Natl. Acad. Sci.} {\bf 97}(18), pp.9840-9843 (2000).

\bibitem{Turanyi} T. Turanyi, A. S. Tomlin and M. J. Pilling, On the error of
the quasi-steady state approximation, {\it J. Phys. Chem.} {\bf 97} 163 (1993)

\bibitem{Kaper2} A. Zagaris, H. G. Kaper and T. J. Kaper, Analysis of the
Computational Singular Perturbation Reduction Method for Chemical Kinetics,
{\it J. Nonlin. Sci.} {\bf 14} pp.59-91 (2004)



\end{thebibliography}
\end{document}